\documentclass[prl,twocolumn,superscriptaddress]{revtex4-1}

\pdfoutput=1

\usepackage{graphicx}
\usepackage{epstopdf}
\usepackage{epsfig}
\usepackage{float}
\usepackage{amsmath}
\usepackage[sort&compress]{natbib}

\allowdisplaybreaks

\begin{document}

\title{Electromagnetically induced transparency in an isotopically purified Nd$^{3+}$:YLiF$_4$ crystal}
\author{Rinat Akhmedzhanov}
\affiliation{Institute of Applied Physics of the Russian Academy of
Sciences, 603950 Nizhny Novgorod, Russia}
\affiliation {Zavoisky Physical-Technical Institute of the Russian
Academy of Sciences, 420029 Kazan, Russia}
\author{Lev Gushchin}
\affiliation{Institute of Applied Physics of the Russian Academy of
Sciences, 603950 Nizhny Novgorod, Russia}
\affiliation {Zavoisky Physical-Technical Institute of the Russian
Academy of Sciences, 420029 Kazan, Russia}
\author{Nikolay Nizov}
\affiliation{Institute of Applied Physics of the Russian Academy of
Sciences, 603950 Nizhny Novgorod, Russia}
\author{Vladimir Nizov}
\affiliation{Institute of Applied Physics of the Russian Academy of
Sciences, 603950 Nizhny Novgorod, Russia}
\author{Dmitry Sobgayda}
\affiliation{Institute of Applied Physics of the Russian Academy of
Sciences, 603950 Nizhny Novgorod, Russia}
\affiliation {Zavoisky Physical-Technical Institute of the Russian
Academy of Sciences, 420029 Kazan, Russia}
\author{Ilya Zelensky}
\affiliation{Institute of Applied Physics of the Russian Academy of
Sciences, 603950 Nizhny Novgorod, Russia}
\affiliation {Zavoisky Physical-Technical Institute of the Russian
Academy of Sciences, 420029 Kazan, Russia}
\author{Alexey Kalachev}
\email{a.a.kalachev@mail.ru}
\affiliation {Zavoisky Physical-Technical Institute of the Russian
Academy of Sciences, 420029 Kazan, Russia}

\date{\today}

\begin{abstract}
We report the first observation of electromagnetically induced
transparency (EIT) in an isotopically purified Nd$^{3+}$:YLiF$_4$
crystal. This crystal demonstrates inhomogeneous broadening of
optical transitions of about 35~MHz. EIT is observed in a
symmetrical $\Lambda$-like system formed by two hyperfine sublevels
of the ground state corresponding to a zero first order Zeeman
(ZEFOZ) transition and a hyperfine sublevel of the excited state,
which is not coupled to other ground-state sublevels. It is found
that transmission of the probe field as a function of the two-photon
detuning demonstrates a comb-like structure that can be attributed
to superhyperfine coupling between Nd$^{3+}$ ions and fluorine
nuclei. The observed structure can be resolved only in the vicinity
of the ZEFOZ point where the homogeneous linewidth of the spin
transition is sufficiently small. The results pave the way for
implementing solid-state quantum memories based on off-resonant
Raman interaction without spectral tailoring of optical transitions.
\end{abstract}

\maketitle




Electromagnetically induced transparency (EIT) is a quantum
interference effect which can be observed in a multilevel atomic
system where resonant absorption is suppressed and a narrow
transparency window is created via coherent interaction with a
strong coupling field~\cite{Harris1997,Fleischhauer2005}. Appearance
of the transparency window is accompanied by strong dispersion and
enhanced nonlinearity, which has led to a variety of applications
including slow light propagation~\cite{Hau1999,Budker1999}, optical
storage~\cite{Phillips2001}, precision
measurements~\cite{Budker2002,Acosta2013}, amplification and lasing
without inversion~\cite{Zibrov1995}, etc. In particular, EIT is a
promising approach to creating optical quantum
memory~\cite{Fleischhauer2000}, which is a basic component required
for scalable quantum computing and long-distance quantum
communication~\cite{Lvovsky2009,Bussieres2013,Heshami2016}. In this
respect, it is important to study EIT in materials promising for
quantum storage. Among them, rare-earth-ion doped crystals at low
temperature have been one of the best candidates~\cite{Thiel2011}
since they provide long coherence times both on optical and spin
transitions, high optical density, no atomic diffusion, possibility
of controlling resonant frequencies and atomic interactions by
external electric and magnetic fields, etc. In particular, EIT has
been extensively studied in the crystals doped by praseodymium ions
~\cite{Ham1997,Ichimura1998,Turukhin2002,Longdell2005,Goldner2009,Akhmedzhanov2006,Heinze2013,Schraft2016},
and up to now the largest EIT storage time ($\sim 1$~min) and
efficiency ($\sim 70$\%) have been demonstrated in the
Pr$^{3+}$:Y$_2$SiO$_5$ crystal~\cite{Heinze2013,Schraft2016}. EIT
was also reported at the telecom wavelength around 1.5 $\mu$m in the
same host material doped by erbium ions~\cite{Baldit2010}.

In this work, we present experimental results demonstrating EIT in
an isotopically purified Nd$^{3+}$:YLiF$_4$ crystal. Neodymium doped
crystals are among promising materials for quantum memory
applications since they demonstrate allowed transitions at the
wavelength around 850 nm, which corresponds to transparency windows
both in optical fibres and atmosphere. In addition, this wavelength
is accessible for modern diode lasers and convenient for single
photon detection. Odd neodymium isotopes demonstrate hyperfine
structures with overall splitting in the GHz range, which is
promising for storing broadband photons and microwave photons. They
have long spin coherence times reaching 9 ms at cryogenic
temperatures~\cite{Wolfowicz2015}. On the other hand, it has long
been known that erbium and neodymium ions doped into isotopically
purified YLiF$_4$ crystals, where only the ${}^7$Li isotope is
present, demonstrate very narrow optical transitions ($\sim
10$~MHz)~\cite{Macfarlane1992,Macfarlane1998}. Such small inhomogeneous broadening
proves to be very attractive for implementing off-resonant Raman
quantum memory
protocols~\cite{Moiseev2011,Moiseev2013,Zhang2013,Kalachev2013,Zhang2014},
which has recently made these crystals a subject of active research
~\cite{Zhou2013,Marino2016,Akhmedzhanov2016,Gerasimov2016,Kukharchyk2017}.
Since hyperfine structure of the optical transition is clearly
resolved in the present material, observing EIT needs no special
ensemble preparation such as spectral tailoring with anti-hole
creating and optical pumping. Moreover, even superhyperfine
structure can be resolved in the EIT process, which opens way for
optical addressing nuclear spin states of surrounding ions and for
quantum storage on the superhyperfine states as well.

\begin{figure}[!h]
\centering
  \includegraphics[width=0.48\textwidth]{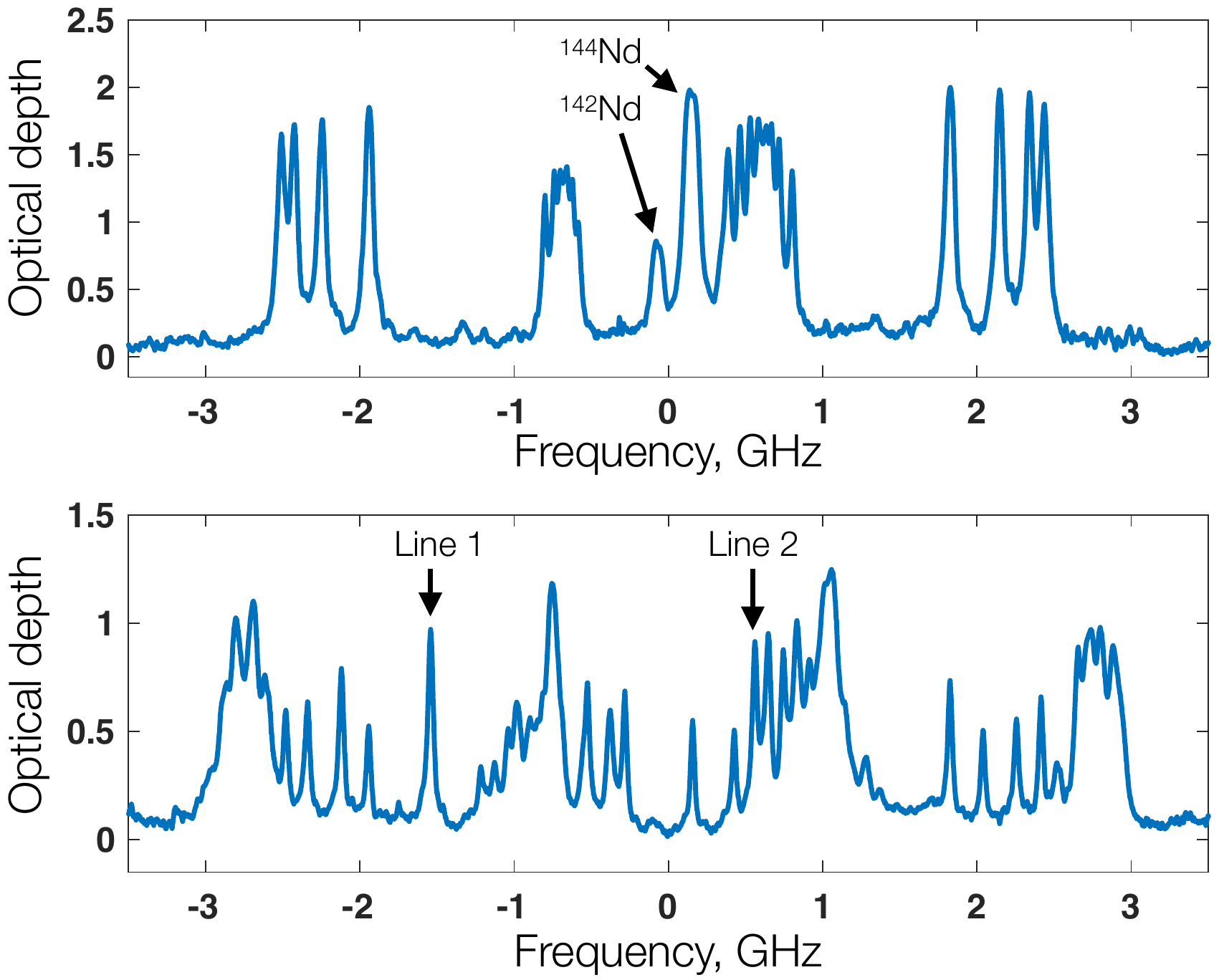}
  \caption{Absorption spectra of the sample on the $^4$I$_{9/2}(1)$--$^4$F$_{3/2}(1)$ transition in the absence of an external magnetic field (above) and in an external longitudinal magnetic field of 60.5~mT (below). The resonant lines 1 and 2 form the $\Lambda$-system that was used for EIT.}
\label{fig1}
\end{figure}

Our experiments were carried out using the Y$^7$LiF$_4$ crystal
(99.7\% ${}^7$Li) containing 0.005 at.\% of $^{143}$Nd$^{3+}$
impurities (96.5\% $^{143}$Nd). The same material was used
in~\cite{Akhmedzhanov2016,Akhmedzhanov2016b} for implementing
quantum memory protocols based on atomic frequency combs. The crystal was grown by using Bridgman-Stockbarger method in an argon atmosphere of high purity (see \cite{Akhmedzhanov2016,Kukharchyk2017} for details). We started
by measuring the absorption spectrum ($\sigma$ polarization) of the
sample on the $^4$I$_{9/2}(1)$--$^4$F$_{3/2}(1)$ transition
(867.5~nm) at 2~K as a function of dc magnetic field directed along
the crystal $c$ axis. Two examples are shown in Fig.~\ref{fig1}.
The isotope $^{143}$Nd has a nuclear spin of $I=7/2$, while the
doubly degenerate electronic states are described by the effective
electron spin $S=1/2$. The hyperfine structure of the ground and
excited states is described by the effective spin Hamiltonian
\begin{align}\label{SpinHam}
H=&g_{\parallel}\mu_B B_z S_z +g_{\perp}\mu_B(B_xS_x+B_yS_y)\nonumber\\&+AI_zS_z+B(I_xS_x+I_yS_y)\nonumber\\&+P[I_z^2-I(I+1)/3],
\end{align}
where $\mu_B=14$~MHz/mT is the Bohr magneton, $g_\parallel$  and
$g_\perp$  are the components of the $g$ factor parallel and
perpendicular to the $c$ axis, $B_i$  are the components of the
external magnetic field, $S_i$ and $I_i$ are the electron- and
nuclear-spin operators, respectively, $A$ and $B$ are the hyperfine
parameters, and $P$ is the quadrupole constant. The ground state
parameters are known from electron paramagnetic resonance (EPR) measurements \cite{Sattler1971} (see
also \cite{Macfarlane1998}): $A=-590$~MHz, $B=-789$~MHz,
$g_{\perp}=2.554$ and $g_{\parallel}=1.987$. The excited state
parameters can be determined from the measured optical spectra as
was done in \cite{Macfarlane1998}, which is posssible due to small
inhomogeneous broadening of the optical transition. In the present work, the resulting
parameters for the excited state are determined to be $A=-257$~MHz,
$B=-456$~MHz, and $g_{\parallel}=0.18$, which are close to the
values obtained in \cite{Macfarlane1998}. In both cases, the
quadrupole constant is found to be negligibly small.

\begin{figure}[!h]
\centering
  \includegraphics[width=0.48\textwidth]{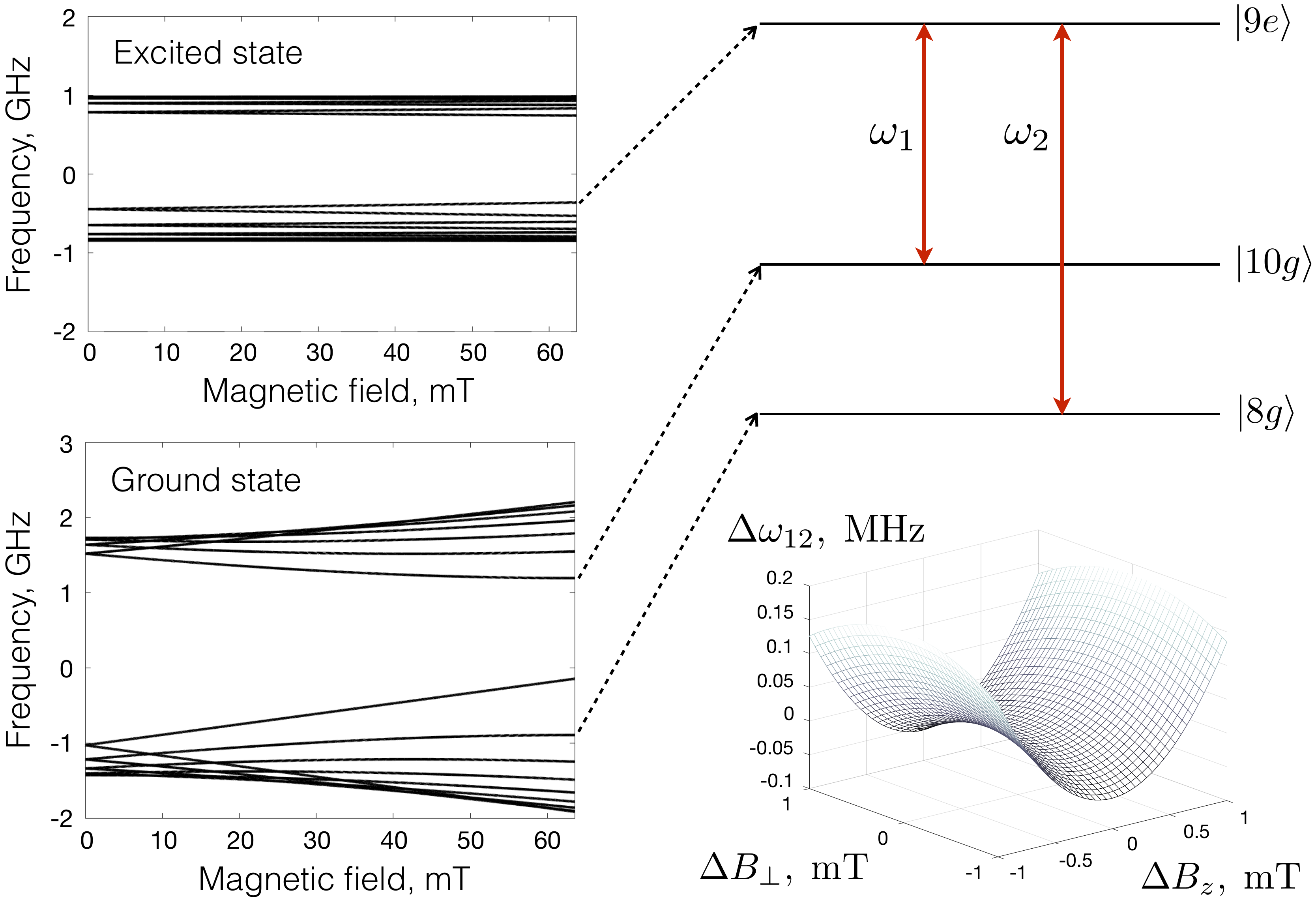}
  \caption{Left: hyperfine structure of the electronic ground and excited states as a function of the longitudinal magnetic field. Right: hyperfine sublevels forming the identified ideal $\Lambda$-system at magnetic field of 63.6 mT (above) and the two-photon transition frequency $\Delta\omega_{12}=\omega_{12}-\omega_{12}^0$ as a function of longitudinal and transverse magnetic field deviation from the ZEFOZ point (below).}
\label{fig2}
\end{figure}

Considering the ground electronic state, for certain values of the
external magnetic field it is expected to observe hyperfine
transitions which are insensitive to the magnetic field
fluctuations. Such zero first order Zeeman (ZEFOZ) transitions
\cite{Fraval2004,Lovric2011,McAuslan2012} can demonstrate extremely long spin
coherence times \cite{Zong2015} thereby providing the optimal
conditions for quantum storage. The calculations show that the most
interesting ZEFOZ transition appears under application of the
longitudinal magnetic field of about 63.6~mT. In this case, the
hyperfine sublevels of the ground state form an ideal symmetric
$\Lambda$-scheme with a hyperfine sublevel of the excited state. To
be more precise, under such a magnetic field we have the following
states:
\begin{align}
|8g\rangle &=\frac{1}{\sqrt{2}}\left(-\left|\frac{5}{2},\frac{1}{2}\right\rangle-\left|\frac{7}{2},-\frac{1}{2}\right\rangle\right),\\
|10g\rangle &=\frac{1}{\sqrt{2}}\left(-\left|\frac{5}{2},\frac{1}{2}\right\rangle+\left|\frac{7}{2},-\frac{1}{2}\right\rangle\right),\\
|9e\rangle &=\left|\frac{7}{2},\frac{1}{2}\right\rangle,
\end{align}
where $|me\rangle$ and $|mg\rangle$ correspond to $m$th sublevel of
the ground ($g$) and excited ($e$) state, respectively, while
$|M_I,M_S\rangle$ stands for the eigenstate of the Hamiltonian
(\ref{SpinHam}) with specific values of the axial component of the
nuclear spin ($-7/2\leq M_I\leq 7/2$) and of the effective electron
spin ($-1/2\leq M_S\leq 1/2$). The excited state sublevel
$|9e\rangle$ is coupled only to the pair of the ground state
sublevels, $|8g\rangle$ and $|10g\rangle$, by equally strong optical
transitions with $\sigma$ polarization (Fig.~\ref{fig2}). The
resulting most efficient and symmetric $\Lambda$-scheme is very
convenient for observing Raman interaction between the atoms and two
optical fields. In particular, numerics show that it provides a large signal-to-noise ratio when single photon states are stored and reconstructed via off-resonant Raman scattering \cite{Berezhnoi2017}. It is worth noting that in the external magnetic
field the inhomogeneous broadening of the optical transitions
involved is reduced to $\sim 35$~MHz compared to $\sim 70$~MHz
observed in the zero field case.
The frequency of the hyperfine transition as a function of magnetic
field around the ZEFOZ point is described by the formula
\begin{equation}\label{omegaAB}
\omega_{12}=\omega_{12}^0+S_{2x}\Delta B_x^2+S_{2y}\Delta B_y^2+S_{2z}\Delta B_z^2    ,
\end{equation}
where $\omega_{12}^0\approx 2087$~MHz is the ZEFOZ transition
frequency, $\Delta B_i$ is the deviation of magnetic field from the
ZEFOZ point, and $S_{2i}=\partial^2\omega_{12}/\partial B_i^2$ are
the curvatures which are calculated to be
$S_{2x}=S_{2y}=-52.7$~kHz/mT$^2$ and $S_{2z}=185.3$~kHz/mT$^2$. The
opposite signs of the second derivatives correspond to a saddle-type
dependence (Fig.~\ref{fig2}). At the same time, the frequencies
$\omega_1$ and $\omega_2$ are linearly proportional to the magnetic
field with the gradient $\partial\omega_{1(2)}/\partial
B_i=1.33$~MHz/mT.

\begin{figure}[!h]
\centering
  \includegraphics[width=0.48\textwidth]{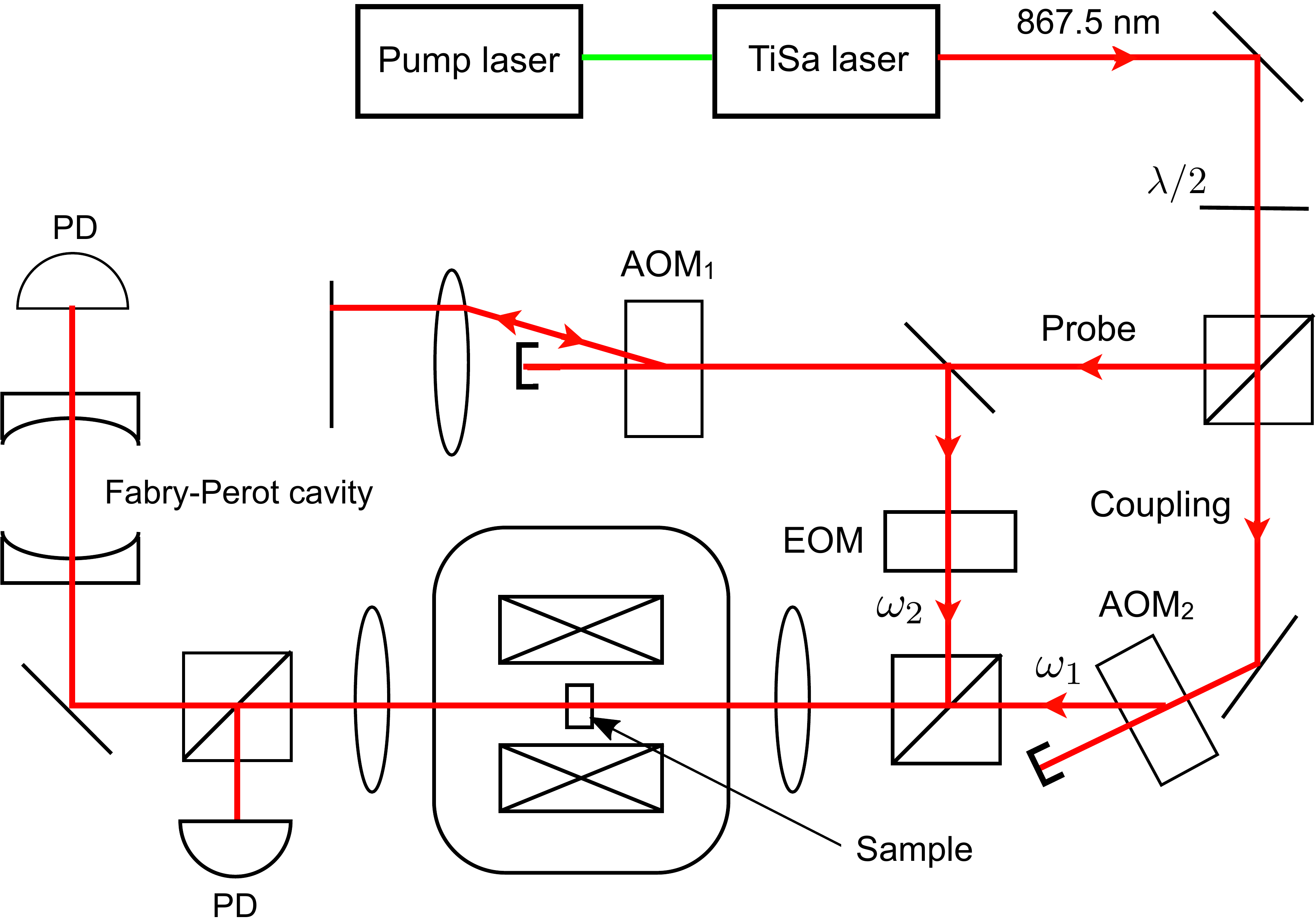}
  \caption{Schematic of the experimental setup.}
\label{fig3}
\end{figure}

The experimental setup for studying electromagnetically induced
transparency is shown in Fig.~\ref{fig3}. The emission of a single
mode Ti:Sa laser (0.5 W power, 500 kHz linewidth) is split into two
separate beams using a polarizing beam splitter. One of the beams is
used as pump, the other as probe. The relative intensities of the
two beams can be adjusted using a halfwave plate placed in front of
the beam splitter. In order to implement the required pulse
sequences and frequency scans, each beam is then passed through an
acousto optic modulator (AOM). The pump beam AOM (AOM$_2$) is set up in a
single pass scheme while the probe AOM (AOM$_1$) is set up in the double pass
scheme to avoid the beam walk during the frequency scan. The probe
beam is then passed through an electro optic modulator (EOM) to
create the required frequency detuning from the coupling beam. The two
beams are then combined on the beam splitter and focused onto the
crystal of 8 mm length that is placed inside a superconducting solenoid. The
crystal and the solenoid are kept at 2~K inside a liquid helium
cryostat. Both laser fields propagate along the optical $c$-axis and
are polarized perpendicular to it ($\sigma$ polarization). After the cryostat the
coupling field is cut off using a polarizer (a Glan prism), while the
probe field is then sent into a tunable confocal
Fabry-Perot cavity for
additional frequency filtering. One of the mirrors is mounted onto a
piezoceramic element so that the cavity length can be controlled by
external voltage. The free spectral range of the cavity was 9.4~GHz
with the finesse of 11.

The described setup is meant to accomplish two tasks. The first is to control how well the coupling and probe beams spatially overlap. This is done by using the coupling beam to burn a long lived spectral hole in the inhomogeneously broadened line which is then read out using the scanned probe beam. The beams are adjusted until the depth of this hole is maximized. For this experiment, the EOM is turned off while the AOM frequencies are tuned so that the central frequencies of the two beams are the same. The second task is to select the appropriate frequency sideband of the probe beam that is modulated by the EOM. This is done by measuring the absorption spectra for different sidebands. Initially, the EOM is turned off and the laser is tuned to the $|10g\rangle - |9e\rangle$ transition (Line 1 in Fig. 1). Then, the EOM is turned on and the frequency is scanned around the expected ZEFOZ transition frequency. We select one of the sidebands using the Fabry-Perot cavity and measure the absorption profile. For the correct sideband, the observed profile corresponds to the part of the spectrum around the $|8g\rangle - |9e\rangle$ transition (Line 2 in Fig. 1). For the wrong sideband, no absorption is observed. In addition, the interferometer also filters out the part of the coupling beam that is not cut off by the polarizer. In experiments the coupling field is tuned to the $|10g\rangle - |9e\rangle$ transition (Line 1) while the probe field is tuned either to the same transition for the hole burning measurements or to the $|8g\rangle - |9e\rangle$ transition (Line 2) for studying the electromagnetically induced transparency. 

\begin{figure}[!h]
\centering
  \includegraphics[width=0.48\textwidth]{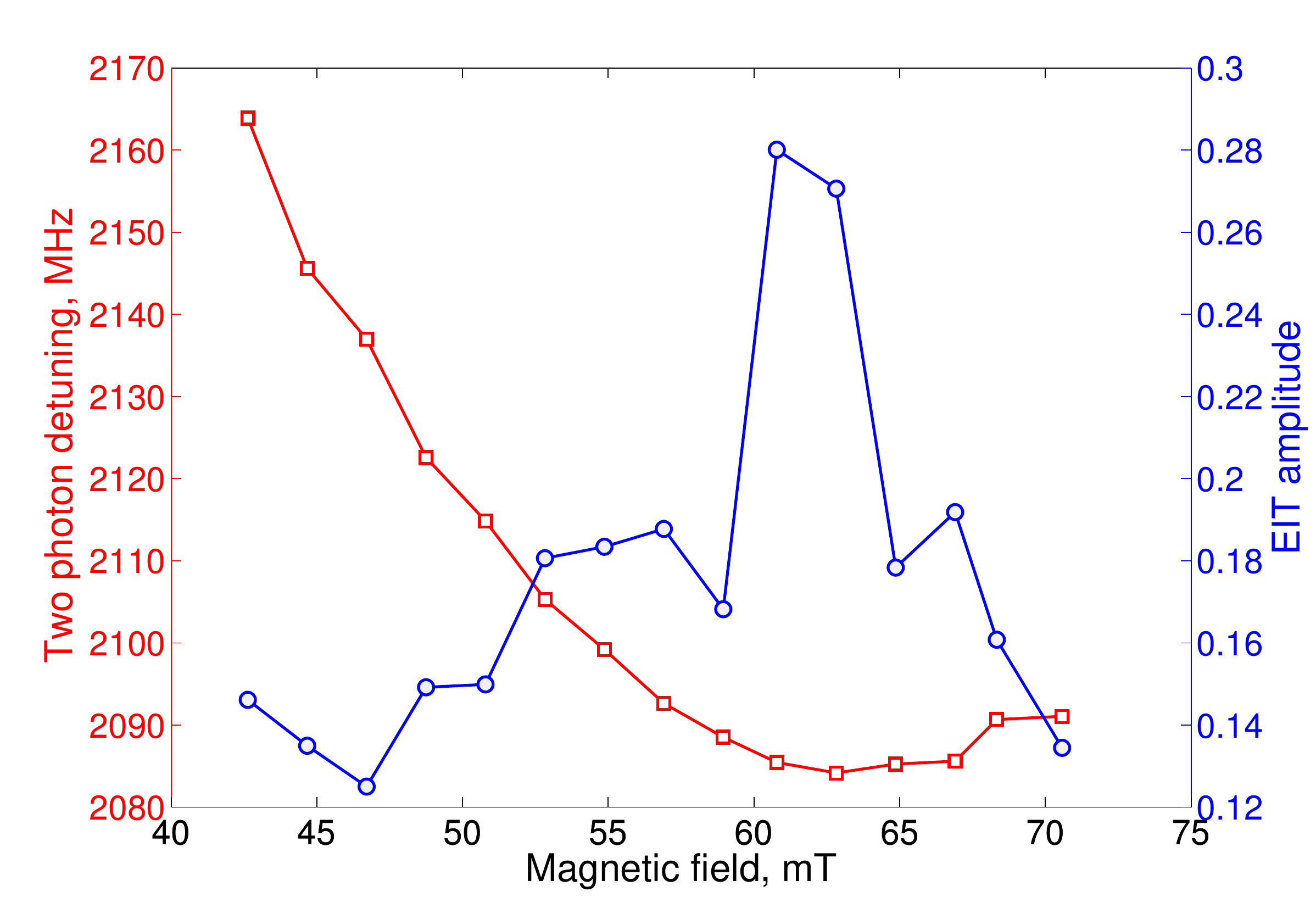}
  \caption{The dependence of the two-photon transition frequency $\omega_{AB}$ (squares) and of the EIT amplitude (circles) on the external magnetic field in the vicinity of the ZEFOZ point. The coupling field intensity is around 50 W/cm$^2$}
\label{fig5}
\end{figure}

EIT can be observed by comparing the intensities of the probe pulse
with and without the coupling field present. When the two-photon
resonance condition is satisfied, the coupling field makes the crystal
more transparent for the probe pulse. The effect is described by the
EIT amplitude which is defined as the ratio $(\alpha_{\rm
off}-\alpha_{\rm on})/\alpha_{\rm off}$, where $\alpha_{\rm on}$
($\alpha_{\rm off}$) is the resonant absorption coefficient with the
coupling field tuned on (off). The experiment is performed for
different values of the external magnetic field. The resulting
values of the EIT amplitude and EIT resonance positions (the values
of the two-photon transition frequency) are shown in
Fig.~\ref{fig5}. It can be seen that the highest transparency
coincides with the location of the ZEFOZ point (the point where the
two-photon transition frequency $\omega_{12}$ attains minimum as a
function of the longitudinal field $B_z$). The dependence of
$\omega_{12}$ on the magnetic field is in good agreement with the
difference between the $|10g\rangle$ and $|8g\rangle$ sublevels
calculated from the Hamiltonian (\ref{SpinHam}).
\begin{figure}[!h]
\centering
  \includegraphics[width=0.48\textwidth]{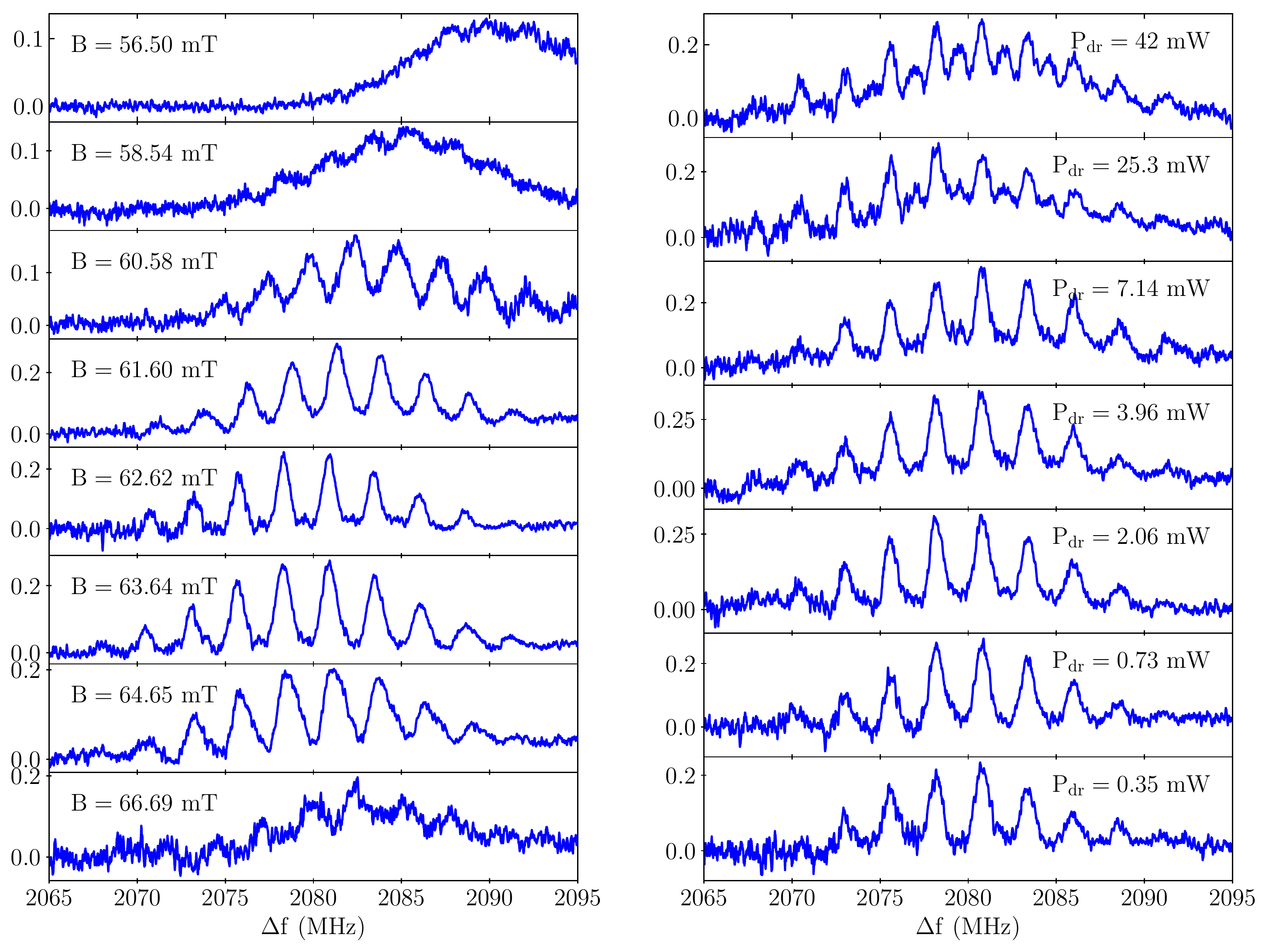}
  \caption{Left: EIT feature profile for different values of external magnetic field. The coupling field intensity is around 7 W/cm$^2$. Right: EIT feature profile for different intensities of the pump field. The crystal is located close to the beam focal point (beam waist around 100 $\mu$m) and the external magnetic field is 63.6~mT. $\Delta f$ is the frequency detuning of the probe field from the coupling field.}
\label{fig7}
\end{figure}

It is worth noting that the total width of the EIT feature is around
12 MHz and it does not depend on the pump intensity or the value of
the external magnetic field. However, under certain conditions the
internal structure of the transparency window changes. For instance,
when the external magnetic field is set to the value corresponding
to the ZEFOZ point, it can be seen that the EIT feature is composed
of nine peaks. The peaks are separated by about 2.8 MHz and each
peak is around 1 MHz wide. When the magnetic field is detuned from
the ZEFOZ point, the width of each peak increases while the distance
between the peaks remains the same (Fig.~\ref{fig7}, left). As a
result, about 7~mT away from the ZEFOZ point the peaks completely
overlap, forming a single 12 MHz wide EIT feature without any
resolvable internal structure. In addition, an interesting effect
can be observed when the pump intensity is increased (see Fig.~\ref{fig7},
right). The resulting high pump intensities cause additional peaks
to appear in the EIT feature. These peaks are located exactly
between the peaks that can be observed at lower intensities.

We believe that the splitting of the EIT feature at low coupling field intensities is caused by superhyperfine interaction between the Nd$^{3+}$ electronic spins and the neighbouring fluorine nuclear spins. Superhyperfine structure in Nd doped YLiF crystals has been previously studied using EPR in external magnetic fields both parallel and perpendicular to the crystal $c$-axis~\cite{Aminov2015}. Similarly to this work, we observe nine separate components in the EIT feature, and the splitting between the peaks is approximately equal to the nuclear Zeeman energy of the fluorine ions (2.5 MHz for the field around 63 mT). The picture is quite similar to what is expected in the case when the superhyperfine interaction is smaller than the Zeeman energy \cite{Aminov2013}. The latter is reasonable because the expectation value of the neodymium magnetic moment in the states $|8g\rangle$ and $|10g\rangle$ is equal to zero. It is worth noting that the hole burned on the $|10g\rangle - |9e\rangle$ transition shows a similar but much less pronounced comb-like structure, which is most likely caused by the superhyperfine splitting of the excited sublevel (cf. \cite{Akhmedzhanov2016}).

While the present experiment demonstrates the power of EIT as a spectroscopy method, the existence of the superhyperfine structure made it impossible to obtain transparencies close to 100\%. High EIT transmission could be achieved by preparing the atomic ensemble in one of the superhyperfine states, which corresponds to polarizing fluorine nuclear spins. It has long been known that the YLiF$_4$ matrix is quite promising for the dynamic polarization of $^7$Li and $^{19}$F nuclei \cite{Antipin1978} so that higher EIT transmission seems feasible. 

The blurring of the comb-like EIT structure with detuning of the
magnetic field from the ZEFOZ point can be explained by increasing
homogeneous broadening of the two-photon transition due to the
magnetic noise. To estimate this effect, we take advantage of a
simple model developed in \cite{Lovric2011} and consider Gaussian
magnetic noise, which allows us to write the linewidth as
\begin{equation}
    \Gamma(\Delta {\bf B})=\Gamma_0+\sum_{i=x,y,z}
        |S_{2i}|(\delta B_i)\sqrt{2(\delta B_i)^2+4(\Delta B_i)^2},
\end{equation}
where $\delta B_i$ is the amplitude (variance) of the magnetic field
fluctuations created by surrounding spins, $\Delta{\bf B}=(\Delta
B_x,\Delta B_y,\Delta B_z)$ is the detuning of the external magnetic
field from the ZEFOZ point, and $\Gamma_0$ is the broadening due to
other effects including non-resolved overlapping of superhyperfine lines
(see~\cite{Aminov2013} for more details).
Then the dependence shown in Fig.~\ref{fig7}, left is simulated
quite well with the curvature values $S_{2i}$ calculated above,
$\Gamma_0 \sim 0.5$~MHz and $\delta B_i \sim 1$~mT. The latter value
proves to be close the local magnetic field estimation obtained in
\cite{Macfarlane1998}. On the other hand, such a line broadening
should be accompanied by the decrease of the EIT amplitude, which
indeed has been clearly observed in our experiment
(Fig.~\ref{fig5}).


To sum up, in this work we observe electromagnetically induced
transparency in an isotopically pure
$^{143}$Nd$^{3+}$:Y$^7$LiF$_4$ crystal demonstrating narrow optical
transitions. We find that when the crystal is placed into an
appropriate external magnetic field corresponding to the ZEFOZ
point, we can resolve narrow peaks in the EIT profile. We believe
that these peaks are caused by the superhyperfine splitting of the
Nd$^{3+}$ hyperfine levels in the ground electronic state. Thus, our
results demonstrate that the crystal is potentially attractive for
implementing Raman schemes of quantum memory involving not only
hyperfine but also superhyperfine states without spectral tailoring
of optical transitions.

The authors thank S.L. Korableva for the crystal growth and Pavel
Bushev for useful comments and discussions. The work was supported
by the Russian Science Foundation, Grant No. 14-12-00806.


\end{document}